\documentclass[10pt,conference]{IEEEtran}
\usepackage{cite}
\usepackage{amsmath,amssymb,amsfonts}
\usepackage{pifont,float}
\usepackage{algorithmic}
\usepackage{graphicx}
\usepackage{textcomp}
\usepackage{xcolor}
\usepackage[hyphens]{url}
\usepackage{fancyhdr}
\usepackage{hyperref}
\usepackage{multirow}
\usepackage{tablefootnote}
\usepackage{array}

\title{Instruction Scheduling in the Saturn Vector Unit}

\def\hpcacameraready{} 

\newcommand\hpcaauthors{Jerry Zhao, Daniel Grubb, Miles Rusch, Tianrui Wei,
  Kevin Anderson, Borivoje Nikolic, Krste Asanovic}
\newcommand\hpcaaffiliation{Department of Electrical Engineering and Computer Science, The University of California, Berkeley}
\newcommand\hpcaemail{ \texttt{\{jzh|dpgrubb|miles.rusch|tianruiwei|kevinand|bora|krste\}@berkeley.edu}}

\author{
  \ifdefined\hpcacameraready
    \IEEEauthorblockN{\hpcaauthors{}}
      \IEEEauthorblockA{
        \hpcaaffiliation{} \\
        \hpcaemail{}
      }
  \else
    \IEEEauthorblockN{\normalsize{HPCA \hpcayear{} Submission
      \textbf{\#\hpcasubmissionnumber{}}} \\
      \IEEEauthorblockA{
        Confidential Draft \\
        Do NOT Distribute!!
      }
    }
  \fi 
}

\fancypagestyle{camerareadyfirstpage}{%
  \fancyhead{}

  \fancyfoot[C]{}
}
\begin{document}
\maketitle

\ifdefined\hpcacameraready 
  \thispagestyle{camerareadyfirstpage}
  \pagestyle{empty}
\else
  \thispagestyle{plain}
  \pagestyle{plain}
\fi

\newcommand{\hpcaheight}{0mm}
\ifdefined\eaopen
\renewcommand{\hpcaheight}{12mm}
\fi


\newcommand{\JZ}[1]{\textcolor{red}{JZ: #1}}
\newcommand{\REV}[1]{\textcolor{red}{#1}}
\newcommand{\DG}[1]{\textcolor{blue}{DG: #1}}
\newcommand{\TW}[1]{\textcolor{teal}{TW: #1}}
\newcommand{\MR}[1]{\textcolor{orange}{MR: #1}}
\newcommand{\cmark}{\ding{51}}%
\newcommand{\xmark}{\ding{55}}%

\begin{abstract}
While the challenges and solutions for efficient execution of scalable vector ISAs on long-vector-length microarchitectures have been well established, not all of these solutions are suitable for short-vector-length implementations. 
This work proposes a novel microarchitecture for instruction sequencing in vector units with short architectural vector lengths.
The proposed microarchitecture supports fine-granularity chaining, multi-issue out-of-order execution, zero dead-time, and run-ahead memory accesses with low area or complexity costs.
We present the Saturn Vector Unit, a RTL implementation of a RVV vector unit.
With our instruction scheduling mechanism, Saturn exhibits comparable or superior power, performance, and area characteristics compared to state-of-the-art long-vector and short-vector implementations.
\end{abstract}

\section{Introduction}

Scalable vector architectures have proven to be a robust instruction-set paradigm for executing modern data-parallel applications across a wide range of microarchitecture design points. 
The heritage of these vector architectures points back to the original Cray-style\cite{cray1} vector machines, which featured very long multi-kilobit architectural vector lengths with many-lane microarchitectures. These ``long-vector'' machines continue to see widespread deployment in scientific computing and machine-learning scenarios, where application vector lengths are suitably long. 

However, long-vector units are seldom deployed where short application vector lengths are more prevalent.
Mobile SoCs require performant execution of audio, video, and DSP workloads that operate on diverse application vector lengths~\cite{swan, ramirez2020risc}, while many embedded applications rely on small matrix operations~\cite{frison2018blasfeo}.
When executing these workloads, large, high-capacity register files provide little performance benefit, as much of the hardware vector length cannot be used.



Even for domains where the application vector lengths tend to be suitably long, power and area constraints further restrict the feasibility of long-vector microarchitectures with many-kilobyte SRAM-backed vector register files.
For these reasons, commercial high-performance mobile and edge SoCs generally integrate compact packed-SIMD (P-SIMD) processors instead of long-vector machines.
These P-SIMD extensions are typically less rich than modern scalable vector ISAs, but can be integrated efficiently due to their simplicity and compactness.
Aggressive instructions scheduling microarchitectures along with precisely tuned kernels are necessary for achieving high utilization in such systems.

We propose a short-vector microarchitecture to address the requirements for compact, efficient data-parallel processors executing modern scalable vector ISAs.
Our vector microarchitecture supports all the architectural requirements of scalable vector ISAs while maintaining high utilization of SIMD functional units and memory bandwidth.
Unlike prior vector microarchitectures, Saturn does not rely on costly capabilities like high-throughput instruction fetch, general out-of-order issue, register renaming, or long architectural vector lengths.

Saturn's vector instruction sequencing microarchitecture supports fine-granularity vector chaining, instruction queuing, and limited multi-issue out-of-order execution with zero dead time.
The sequencing mechanism requires minimal hardware resources to implement and fits neatly within a compact decoupled vector execution backend.

We develop and open-source Saturn, a RTL implementation of our microarchitecture implementing the full RISC-V vector extension 1.0 (RVV), a representative example of a modern scalable vector ISA.
Saturn supports the complete RVV 1.0 ISA, including complex addressing modes, memory translation, and precise traps.

Using Saturn, we perform a simulation and VLSI-driven evaluation of the power, performance, and area characteristics of our proposed microarchitecture.
We compare against baseline long-vector and short-vector microarchitectures.
We further evaluate a range of design space parameters exposed by our microarchitecture, specifically around sensitivity to issue queue depth, chime length, memory latency, and application vector length.

We summarize our contributions below:
\begin{itemize}
    \item Saturn, a complete RVV 1.0-compliant short-vector microarchitecture with a physical performance, power, and area evaluation. 
    \item A vector instruction sequencing mechanism that maximizes datapath utilization with short vector lengths.
    \item A comparison of Saturn against baseline long-vector and short-vector designs.
    \item A discussion on the implications of critical design parameters in Saturn's instruction scheduling microarchitecture
\end{itemize}

Our evaluation demonstrates that combining \textbf{short hardware-vector-lengths}, \textbf{compact SIMD datapaths}, and \textbf{dynamic vector instruction scheduling} with a modern \textbf{scalable vector ISA} yields a performant, efficient, and programmable microarchitecture for extracting vectorized DLP.
\section{Background}

We discuss specific challenges of implementing modern scalable vector ISAs, compare the long-vector and short-vector archetypes, and assert the advantages of short-vector units.

\subsection{Scalable Vector Architectures}

The growing complexity of modern data-parallel workloads has reinforced the importance of binary compatibility and performance portability across varied design points.
This has prompted a movement away from traditional P-SIMD ISAs towards modern \textit{scalable vector architectures}~\cite{next_gen_vector}.
Chief among these are the ARM Scalable Vector Extensions (SVE)\cite{arm_sve}, M-Profile Vector Extensions (MVE)~\cite{arm_mve} and RISC-V Vector Extensions (RVV)~\cite{vector_extension}.

While the machine vector length (MVL) is a static architectural constant in SIMD ISAs, in scalable vector ISAs the MVL is an implementation-defined parameter made discoverable to software.
Applications written for a scalable vector ISA can be expressed as MVL-agnostic stripmine loops.
This ensures binary compatibility with some expectation of performance portability across a wide range of potential implementations.

We identify three specific properties of modern scalable vector ISAs that substantially differentiate the backend microarchitecture of a vector implementation from that of a P-SIMD implementation.

\subsubsection{Variable Chime Lengths} 
The \textit{chime length} in a data-parallel microarchitecture can be defined as the ratio of the architectural register width (VLEN) to the datapath width (DLEN)~\cite{chime_length}.
One design challenge of modern vector ISAs is the imposition of variable dynamic chime lengths during program execution.
This arises from the observation that the number of architectural registers needed by a vector kernel to avoid register spilling is highly varied.
Simple, yet common kernels, like \texttt{memcpy} or \texttt{saxpy}, may only need a handful of registers, while complex or unrolled kernels may need tens of registers.

To maximize register file utilization and to reduce instruction fetch pressure across all these kernels, modern scalable vector ISAs have adopted the concept of \textit{register grouping}.
With register grouping, a single instruction can be specified to encode operations over a group of vector registers, effectively treating the group as a single longer vector register.
RVV facilitates register grouping with a length multiplier (LMUL) control field,  while ARM SVE and ARM MVE directly provide \textit{multi-vector} instruction encodings.

Since software will frequently leverage register grouping to reduce dynamic instruction counts, implementations should not expose a performance overhead when executing register-grouped instructions, compared to an equivalent block of single-vector instructions.
This effectively requires that implementations remain efficient at varying chime lengths, since register grouping directly scales the chime length.

\subsubsection{Irregular Vector Memory Instructions}
To enable efficient execution of kernels where vector operand data is interleaved in memory, modern vector ISAs feature load and store instructions that directly de-interleave memory into vector registers.
RVV provides ``segmented" load and store instructions, while SVE provides ``structure" loads and stores.

Since execution of these instructions should still fully utilize the available memory bandwidth, vector microarchitectures must perform a streaming transpose of memory data into vector registers.
Although complex to implement, these ``transpose" or ``segmentation" operations should also not stall the pipeline or disallow chaining.

\subsubsection{Decoupled Implementation Support}
Packed-SIMD implementations have historically been designed as extensions within the existing scalar pipeline of general-purpose cores to reuse existing scalar functionality, while modern vector ISAs are designed to more easily support decoupled implementations.
A rich set of vector instructions, especially the mask instructions, reduce the need for introducing scalar-vector data or control dependencies into vector loops.

RVV avoids overlaying the vector registers on scalar registers to simplify decoupling.
SVE preserves the overlay, but the Streaming SVE~\cite{armv9} variant, which explicitly targets decoupled implementations, uses a different scalar register space than the conventional scalar floating-point register space.

Since modern vector ISAs require precise resumable traps and scalar-vector memory ordering, a modern vector microarchitecture exhibits characteristics of both tightly-integrated SIMD implementations and classic decoupled long-vector implementations.
Specifically, precise exceptions must be reported ahead of commit, deep in the scalar pipeline.
However, the execution backend could still be implemented as a decoupled post-commit unit without degrading performance.


\subsection{Long-Vector Machines}

We characterize \textit{long-vector machines} as those implementations bearing close resemblance to the original Cray vector units.
Modern long-vector machines feature banked, lane-distributed vector register files, typically implemented as dense SRAM macros.

The dense register file in these designs enables very long machine vector lengths, spanning into the multi-kilobit range.
The long vector lengths imply long chime lengths and deep temporal execution, which simplify hazard resolution and reduce the dead-time penalty.
As long as the instruction sequencing mechanism determines an efficient execution plan that maximizes utilization and avoids data and structural hazards, low instruction throughput or dead-time stalls are amortized away by the long chime length.


Examples of modern decoupled long-vector implementations include Ara~\cite{ara2}, Hwacha~\cite{hwacha_manual}, Vitruvius~\cite{minervini_vitruvius_2023}, and the NEC-Aurora~\cite{nec_aurora}.
All these implementations feature lane-distributed SRAM-backed vector register files, and represent the state of the art for compute-maximizing long-vector microarchitectures.

To summarize, long vector microarchitectures rely on \textbf{low-instruction-throughput sequencing of highly-efficient execution plans with long vector chimes} to maximize datapath utilization.

\subsection{Short Vector Machines}

In contrast to long-vector machines, short-vector machines feature a unified multi-ported register file driving a unified cluster of SIMD functional units.
At a high level, these machines are more similar in datapath structure and register-file organization to traditional packed-SIMD machines, but must still handle the complexities of executing a scalable vector ISA.


Shorter vector lengths also dictate the need to execute shorter chimes with high efficiency, precise chaining, and zero dead time.
Notably, the cost of handling short chime lengths in vector implementations increases as the chime length approaches 1, since the required instruction throughput to saturate a vector functional unit is inversely proportional to chime length.
Thus, short-vector implementations, with shorter chime lengths, must sequence instructions with higher throughput than an implementation with longer chime lengths.

Additionally, portable vector workloads likely contain vector code that is suboptimally scheduled for the target vector implementation.
To ensure efficient execution in spite of suboptimal code, short-vector implementations should support some degree of flexible out-of-order instruction scheduling, with chaining and simultaneous issues across the load, store, and execute paths.
These combined requirements pose unique challenges for the sequencing microarchitecture of a short-vector implementation.

To summarize, unlike P-SIMD or long-vector microarchitectures, a short-vector design needs to maximize datapath utilization through \textbf{efficient, precise sequencing of short vector chimes without relying on high instruction throughput or complex instruction scheduling}.

\subsection{Why Short Vectors}

We identify three key advantages of short-vector microarchitectures over long-vector microarchitectures.

\textbf{Many application domains feature short application vector lengths and/or narrow datatypes}, too short to fully utilize a vector register in a long-vector microarchitecture~\cite{ramirez2020risc}.
For instance, Khadem et. al,~\cite{swan} found that the performance of many mobile workloads exhibit sublinear, or even inverse scaling with SIMD widths above 256 bits.
Frison et. al,~\cite{frison2018blasfeo} found that embedded numerical optimization fundamentally relies on small matrix sizes.

Such small problem sizes would map inefficiently to a long-vector microarchitecture with a long native hardware vector length, as much of the vector register storage would be unused.
Critically for performance, long-vector microarchitectures designed around temporal execution over long vector lengths may be less efficient with short vector lengths.
Conversely, short-vector microarchitectures seek to achieve high datapath utilization even with short application vector lengths using their precise sequencing mechanisms.

\textbf{Even with long application vector lengths, short vectors microarchitectures still present some inherent advantages over long-vector units.}
A lower-capacity VRF has positive implications for area and power.
Additionally, short hardware vector lengths do not imply low performance with long application vector lengths; a short-vector machine can be as performant as a long-vector machine, but with superior power and area characteristics.
Notably, the register grouping feature in modern scalable vector ISAs directly provides a mechanism through which a short-vector microarchitecture can execute longer vector lengths with minimal area or power overhead.

\textbf{Short-vector microarchitectures share physical characteristics with existing P-SIMD datapaths commonly deployed in commercial SoCs}, reducing the barrier to deployment for such designs.
While long-vector microarchitectures are not deployed in any commercial mobile or desktop SoCs, compact P-SIMD cores are pervasive, and have not been displaced in mobile or edge domains by specialized accelerators.
For instance, the HVX P-SIMD extension and datapath in the Hexagon DSP processor is critical to compute-intensive tasks on all Snapdragon mobile SoCs~\cite{codrescu2014hexagon,codrescu2015architecture}.
Tens of billions of ARM cores with NEON, SVE, or MVE extensions ship annually to meet this market~\cite{armshareholder}.

Short-vector microarchitectures have the potential to provide the same necessary capability for efficient data-parallel compute in a compact and efficient physical footprint as these P-SIMD systems, but while executing a more future-proof and programmer-friendly scalable vector ISA.



\section{Short-Vector Microarchitecture}

\begin{figure}[h]
\centering
\includegraphics[width=0.90\linewidth]{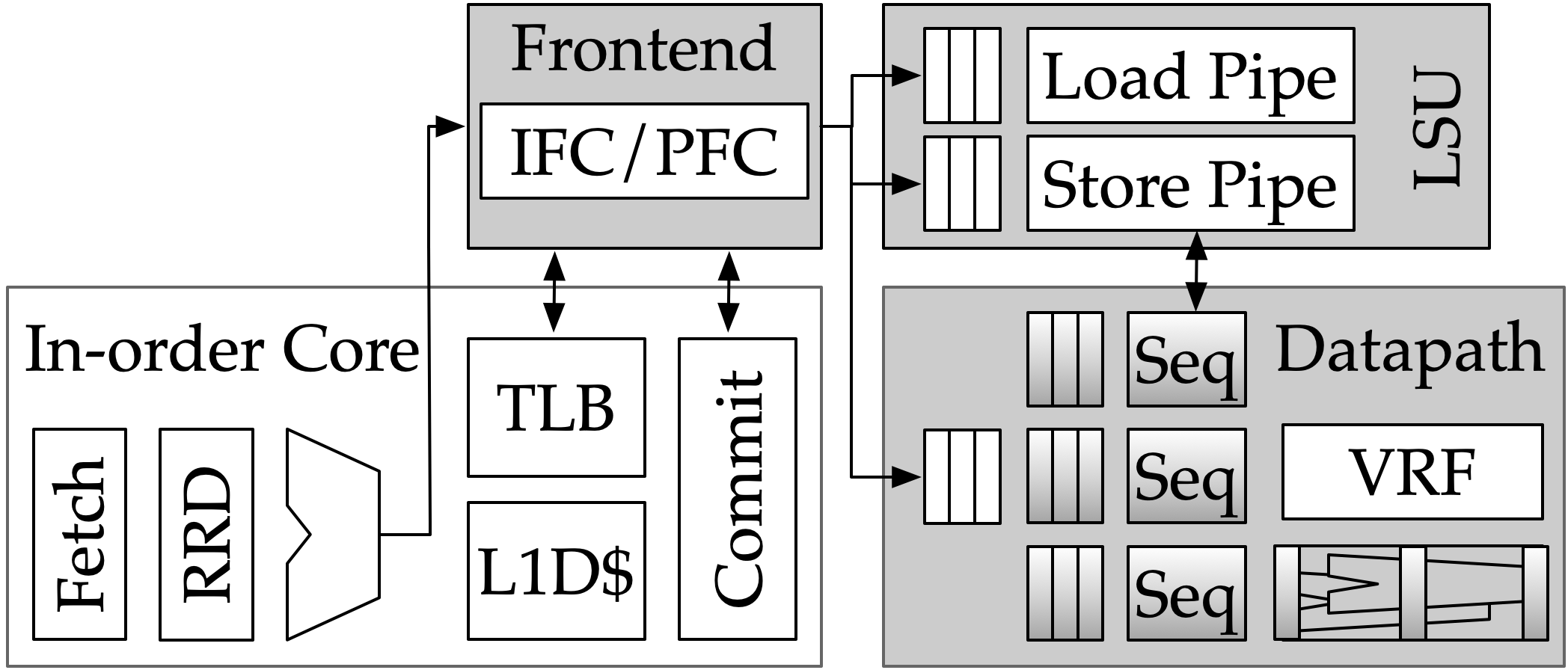}
\caption{Overview of the Saturn short-vector microarchitecture (gray) and its intended integration into an in-order core. The gradated regions indicates components relevant to the scheduling mechanism.}
\label{fig:overview}
\end{figure}

\begin{figure}[h]
\centering
\includegraphics[width=0.85\linewidth]{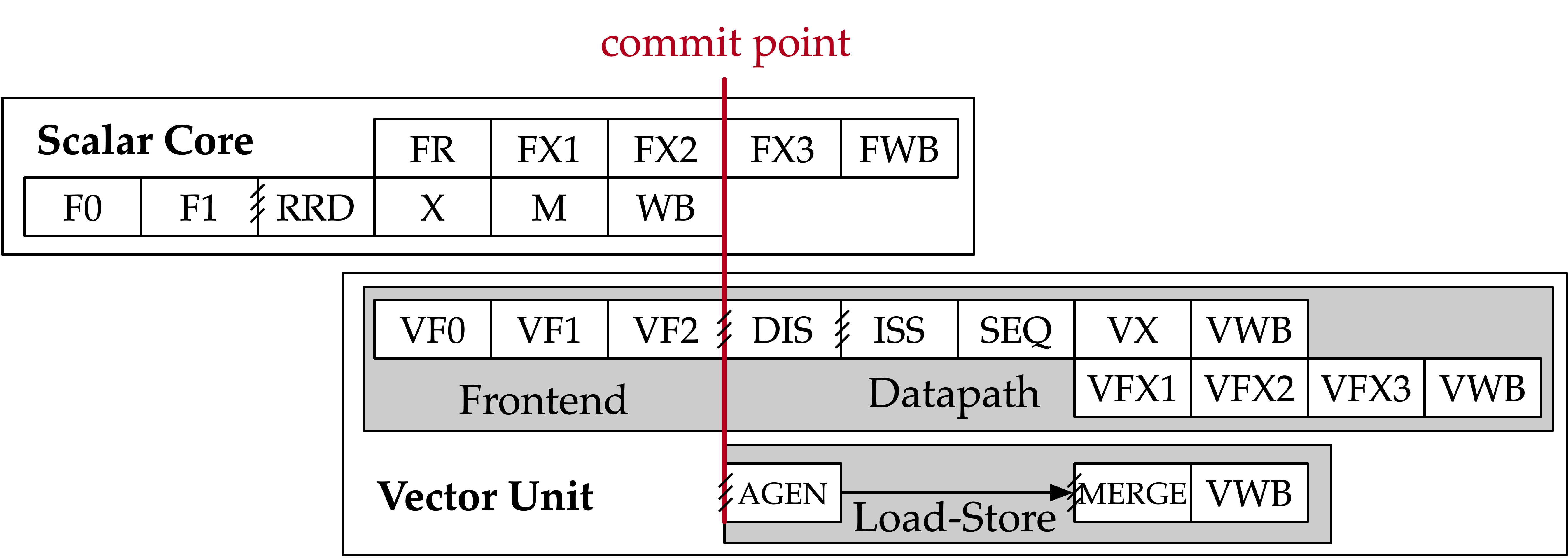}
\caption{Pipeline stages of Saturn when integrated into a host in-order RISC-V core. Hatched lines indicate queues between stages.}
\label{fig:pipe}
\end{figure}

Figure \ref{fig:overview} depicts how all three components of Saturn integrate into a minimal host in-order CPU.
In this section we describe how these components can be microarchitected to support a modern vector ISA with short vector lengths.
We provide only a cursory overview of these components here.
A more detailed description is available at \href{https://www.saturn-vectors.org}{saturn-vectors.org}.
\begin{itemize}
\item The \textbf{Frontend} sits ahead of the commit point of the host CPU and implements checks for precise vector faults in a shallow pipeline.
\item The \textbf{Load/Store Unit} receives post-commit vector instructions and issues requests into the memory system.
\item The \textbf{Backend Datapath} includes the vector register file, instruction sequencers, and functional units
\end{itemize}

\subsection{Frontend}

The frontend precisely checks for the presence of faults in vector memory instructions and enables decouplng of the post-commit load-store and execute datapaths.
Unlike P-SIMD microarchitectures, this vector frontend does not reuse the scalar load/store unit.
The frontend merely checks for potential faults, while the actual address-generation and memory-issue occurs in the separate post-commit load-store unit.

The frontend operates in two modes: \textit{pipelined} or \textit{iterative}.

\subsubsection{Pipelined}
The pipelined mode executes in lock-step with the scalar core's pipeline.
This mode computes a conservative bound on the extent of the vector memory access from the base address provided by a scalar register, and the encoded vector register length and addressing mode.
For unit-strided or constant-strided accesses, the bound computation is trivial and can be performed in the same stage as scalar address-generation.

Single-page contiguous-accesses to a homogeneous physical memory region can be verified to be free of fault with a single TLB access.
Such instructions can be safely committed and issued to the backend with the physical address without stalling any instructions.
Multi-page contiguous-accesses are cracked into single-page operations, while indexed or strided accesses are deferred to the iterative mode.
In summary, the pipelined mode is designed to catch the common-case contiguous unit-stride acceses that must be issued to the load-store unit with high throughput.

\subsubsection{Iterative}
This mode prevents the scalar core from retiring instructions younger than the currently processed vector instruction, giving cycles for an iterative state machine to step through the instruction element-wise, fetching indices from the backend, and precisely computing the accessed address for each element.
While this procedure imposes a very high performance overhead, it is used only to support high-extent indexed memory operations which would fundamentally be performance-limited by a single-ported TLB, TLB reach, and page-table walks.

\begin{figure}[t]
\centering
\includegraphics[width=0.95\linewidth]{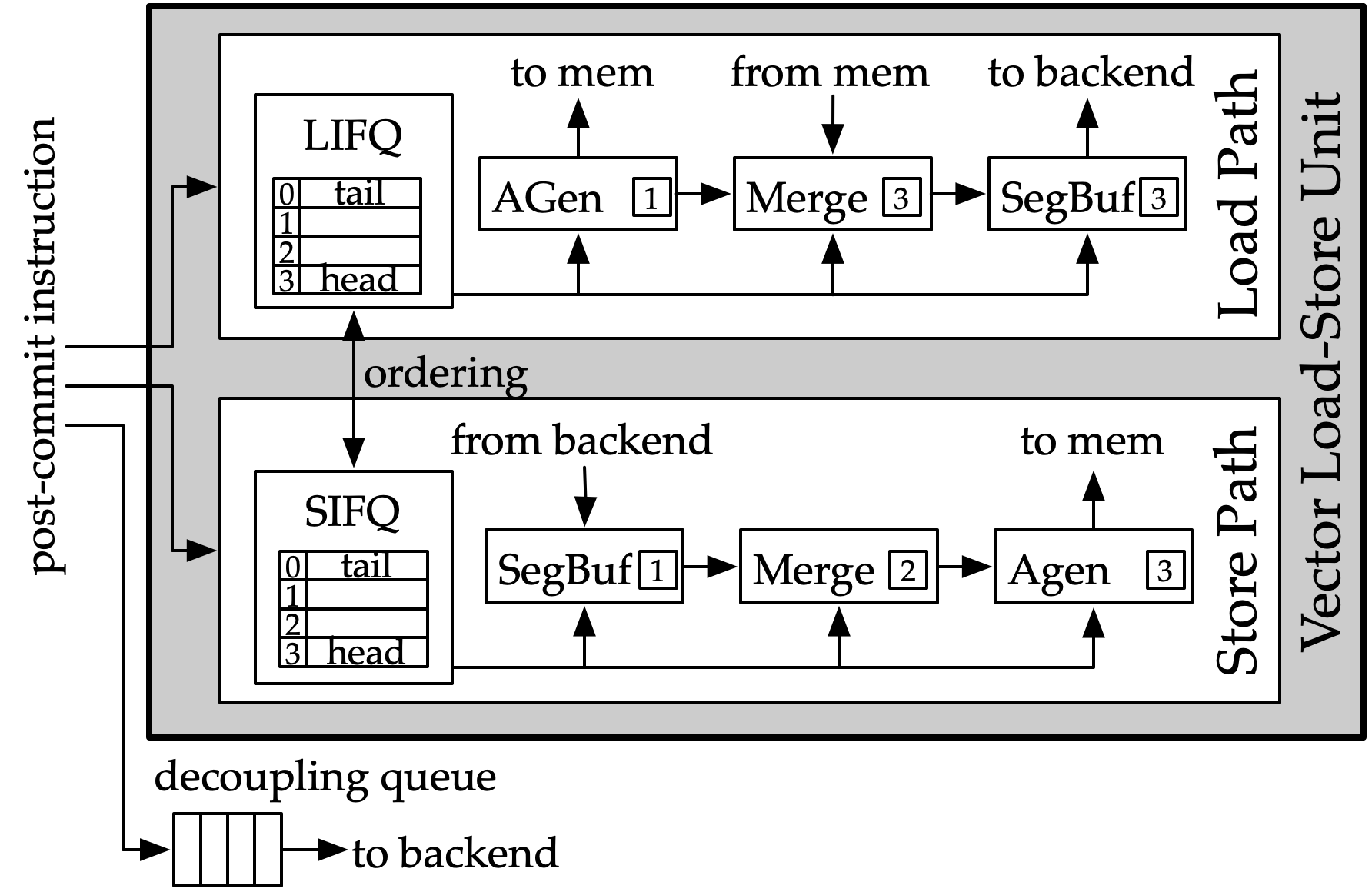}
\caption{The vector load and store paths handle variable-chime and long-latency memory operations with minimal storage requirements. The load path depicts a long-latency-load in the Agen unit running ahead of a long-chime load in the merge and SegBuf units. The store path depicts a sequence of short-chime stores in the SegBuf, Merge, and Agen units.}
\label{fig:lsu}
\end{figure}
\subsection{Load-store unit}

The vector load-store unit is microarchitecturally similar to per-lane load-store paths in long-vector implementations and load-store paths in decoupled SIMD microarchitectures.
In Saturn, this unit is implemented like the "access processor" in the decoupled-access-execute (DAE) paradigm~\cite{decoupled_execute} instead of an integrated memory pipeline.

Following the DAE paradigm allows exploiting memory-level parallelism by adjusting the size of low-cost decoupling queues without further costly implications on the backend datapath and control components.
Figure \ref{fig:lsu} depicts the independent load and store paths within the load-store unit.
We provide only a broad overview of the rest of the load-store unit components here.

Load and store in-flight queues track the base and extent of in-flight or pending vector memory operations. 
Vector instructions are tracked in the in-flight queues without requiring cracking, minimizing the structural cost of long-chime instructions.
Address CAMs across each queue are used to enforce scalar-vector and vector-vector memory disambiguation.

Separate load and store paths enable simultaneous issue of loads and stores.
Both paths can be designed as pipelined streams of latency-insensitive units.
Each path processes memory requests and packets in-order, and are driven by pointers into the circular in-flight-load and in-flight-store queues.

Merge units correct for misaligned memory accesses and convert between sparse memory packets and dense vector register rows.
Segment buffers provide streaming, high-throughput reformatting of segmented fields into vector registers.

The load path is decoupled from the rest of the backend, enabling run-ahead load address generation to exploit memory-level-parallelism across high memory latency memory systems.
Similarly, the store path runs behind the backend.

\subsection{Backend Datapath}

\begin{figure}[t]
\centering
\includegraphics[width=\linewidth]{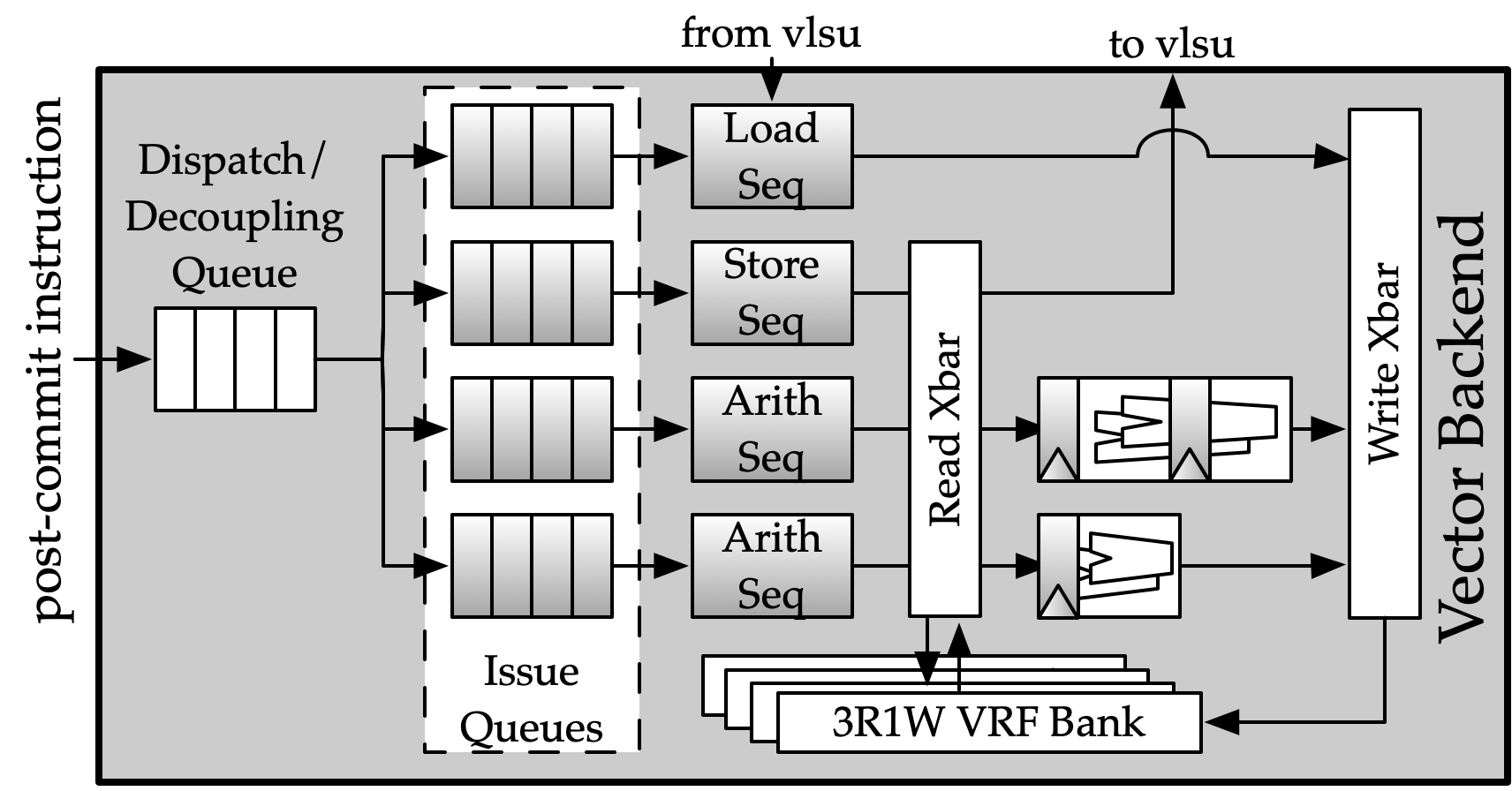}
\caption{The backend organization for a configuration with two arithmetic sequencers, separate load/store sequencers, 4-entry instruction queues, and a 4x3R1W vector register file. Gradated regions indicate the out-of-order execution window.}
\label{fig:backend}
\end{figure}

The backend datapath comprises the issue queues, instruction sequencers, unified vector register file (VRF), and SIMD execution units.  
The issue queues buffer vector instructions and feed the sequencers, which crack vector instructions into single-cycle micro-ops of execution.
Upon issue, vector micro-ops read their operands from the VRF before proceeding through the SIMD execution units.


The backend is organized around ``element-groups'' as the base unit of compute, where an element-group is a DLEN-wide segment of a vector register.
The width of each VRF bank, the width of the register access crossbars, and the width of the SIMD functional units are all DLEN.
Thus, an element-group represents a packet of register file data that can be consumed or produced in one cycle.

The VRF is implemented as a banked multi-ported flip-flop array.
As shown in Figure \ref{fig:backend}, read and write crossbars arbitrate for access to the VRF ports across the banks.
The vector register file is striped across the banks by element group; neighboring element groups reside in consecutive banks.

The sequencers each execute independently, enabling dynamic ``slip" across the load, store, and arithmetic paths.
This yields limited out-of-order execution of vector instructions. 
Combined with the decoupling queues and fine-grained hazard tracking, this sequencer structure can tolerate many suboptimal code patterns through dynamic sequencing.

The entire backend pipeline can be implemented with very few pipeline stages, improving its suitability for compact implementations.
Aggressive implementations can bypass directly from the dispatch queue into the sequencers.
Our hazard checking also requires minimal logic depth and can be implemented in the same stage as register-read. 
Thus, an instruction can proceed from the dispatch queue to vector-register-write-back in as few as three cycles.

\section{Instruction Scheduling in Short-Vector Units}

Saturn's instructions scheduling mechanism performs \textbf{vector sequencing} with \textbf{explicit chaining}.
We first motivate these design decisions before discussing the details of the scheduling microarchitecture.

\subsection{Cracked vs Sequenced Scheduling}

\begin{figure}[h]
\centering
\includegraphics[width=\linewidth]{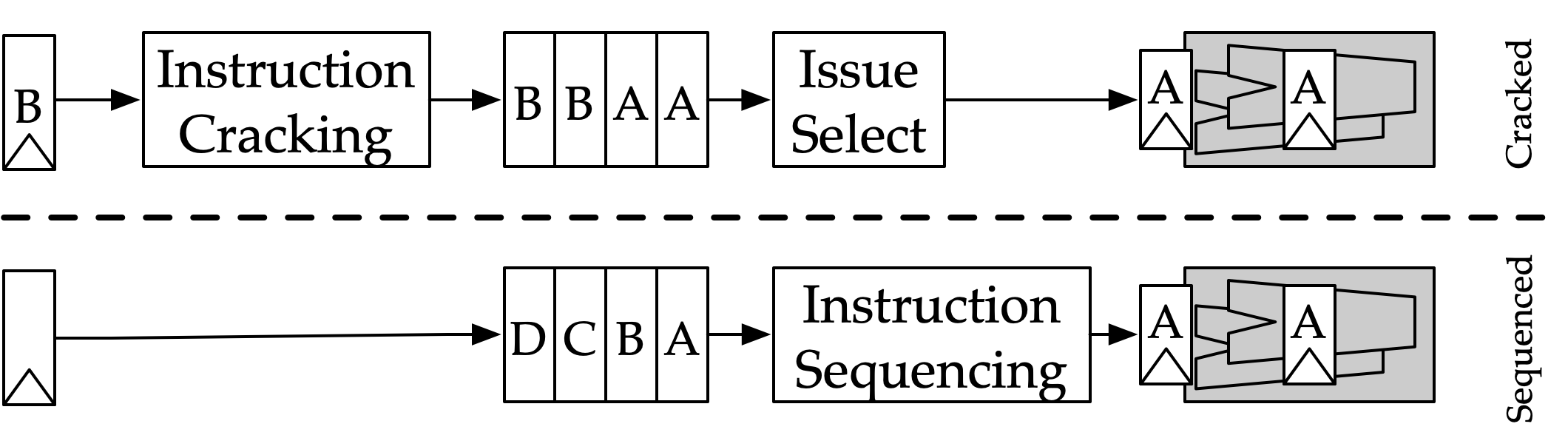}
\caption{A comparison of instruction cracking vs sequencing for a block of vector instructions A/B/C/D. The cracking approach will stall dispatch without deep issue queues.}
\label{fig:datapath}
\end{figure}

One approach to schedule vector instructions is to crack them into single-cycle micro-ops early in the pipeline, ahead of where hazard resolution occurs at issue select.
This approach is especially tempting as a means to handle register-grouped instructions by cracking them into single-vector-register instructions.
Although this approach reduces hazard resolution complexity, it tends to stall dispatch without deep issue queues, as shown in Figure \ref{fig:datapath}.

Vector sequencing cracks vector instructions into micro-ops behind the issue queues.
Since issue queue entries may encode many cycles of work accessing many vector elements or registers, managing hazards across the issue queues is more costly.
However, late sequencing reduces pressure on issue queue depth.

\subsection{Explicit vs Implicit Chaining}

With implicit chaining microarchitectures, the microarchitecture relies on regular vector access patterns to effect chaining.
This approach relies on the observation that if the source and sink instructions both write and read their operands at the same fixed rate, the sink instruction can \textit{always} schedule the next micro-op after a source instruction writes back \textit{any} element.

While this approach is attractive for its low cost and ease of implementation, it struggles to handle irregular vector instructions that produce or consume operands at irregular rates.
It additionally cannot elegantly handle variable-latency memory systems without global stalls.
In contrast, explicitly chained vector units precisely track the availability of operands at sub-register granularity.
Explicit chaining microarchitectures opportunistically perform chaining in response to dynamic behaviors.

\subsection{Vector Instruction Scheduling}

The instruction window for out-of-order execution in Saturn includes the per-sequencer issue queues and the instructions within the sequencers themselves.
Thus, the scheduling mechanism must resolve potential data hazards across all the issue queues, sequencing instructions, and sequenced micro-ops.
We define this set of instructions and micro-ops as the \textit{OoO window}, as shown in Figure \ref{fig:backend}.

To resolve structural hazards on VRF ports and the execution units, the microarchitecture implements read and write port arbitration across all the sequencers.
The sequencers advertise the requested read and write addresses for the next micro-op to the arbiters.
Bank or port conflicts induce a structural hazard and stall the sequencing of the younger instruction.
For data hazards, the sequencing mechanism should track data dependencies at element group granularity and support chaining across RAW, WAW, and WAR conditions.

Resolving data hazards using traditional CDC6600-like scoreboarding\cite{scoreboard} is a poor fit for an explicit chaining system where in-order register read and write-back are not enforced.
We propose an augmented scoreboarding scheme which tracks pending read/write scoreboards (PRSb/PWSb) for each instruction in the issue window at element-group, rather than register granularity.
We divide our discussion into a section on the microarchitectural state required for our algorithm, followed by a discussion of the algorithm itself.

\subsubsection{Tracking Data Hazards} 
Our approach tracks the PRSb and PWSb scoreboards for all instructions within the OoO instruction window.
The bit-width of each is the total number of the element groups in the VRF, or, $(\text{VLEN}\times\text{\# registers})/\text{DLEN}$.

In the general case, this would be an infeasibly costly state to represent across all the necessary components in the OoO window.
However, for short-vector microarchitectures with shallow pipelines and issue-queues, we demonstrate that the structural cost to represent the scoreboards is minimal, or already provided as part of existing state in the backend.
The only additional state is a unique tag to enable age disambiguation between instructions in the OoO window. 
This tag is assigned when an instruction enters the OoO window and freed when the instruction completes sequencing.

\begin{table}[h]
\centering
\caption{Per-instruction Scoreboard Table, for a machine with 4 Vector Registers and VLEN=2, DLEN=2}
\label{tab:sboard}
\renewcommand*{\arraystretch}{1.2}
\begin{tabular}{|l|c|c|c|}
\hline 
\textbf{Instruction} & \textbf{ID/Age} & \parbox{5em}{\centering{\vspace{0.3em}\textbf{PRSb}\vspace{0.3em}}} & \parbox{5em}{\centering{\vspace{0.3em}\textbf{PWSb}\vspace{0.3em}}}
\\ \hline \hline
\texttt{vadd.2 v0, v0, v2}  & 0 & 8'b00000000 & 8'b00001\underline{1}00\\ \hline
\texttt{vle.2  v2}          & 1 & 8'b00000000 & 8'b11\underline{1}00000\\ \hline
\texttt{vadd.2 v0, v0, v2}  & 2 & 8'b111\underline{1}111\underline{1} & 8'b00001111\\ \hline
\texttt{vle.2  v2}          & 3 & 8'b00000000 & 8'b11110000\\ \hline
\texttt{vadd.2 v0, v0, v2}  & 4 & 8'b11111111 & 8'b00001111\\ \hline
\end{tabular}
\end{table}
\begin{figure}[h]
\centering
\includegraphics[width=\linewidth]{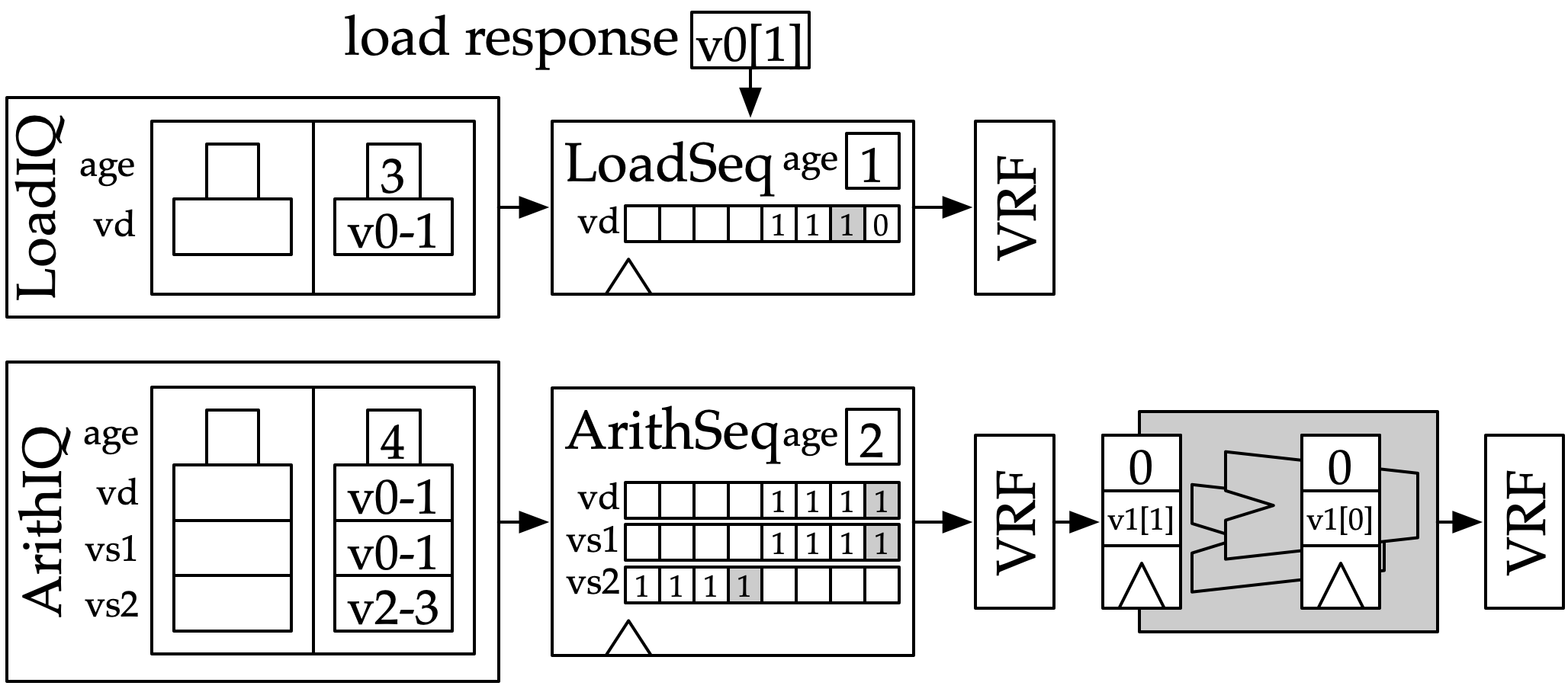}
\caption{A diagram depicting how Saturn implements the PRSb and PWSb scoreboards for instructions in Table \ref{tab:sboard}. At the next clock edge, the load sequencer will sequence a load into register v1[0], while the arithmetic sequencer will sequence reads from v0[0] and v2[0] to effect a pipelined write to v0[0]. Gray sequencer scoreboard elements will be cleared on the next clock edge.}
\label{fig:hazarding}
\end{figure}

Table \ref{tab:sboard} depicts the PRSb and PWSb scoreboards for a simple register-grouped loop with RAW, WAW, and WAR hazards, where all these instructions are currently in the OoO window.
In this example, the machine has four architectural registers, with each register comprised of two element groups. 
The code sequence in this example leverages register grouping on pairs of vector registers.
Instructions 0, 1, and 2 are currently in-flight, while instructions 3 and 4 are resident in the issue queues.
On the next clock edge, instructions 0 and 1 will perform writes, while instruction 2 will perform a read, clearing the underlined bits in Table \ref{tab:sboard}.

Within the issue queues, the PRSb and PWSb need only to be known coarsely since no reads and writes have been performed at this stage.
Thus, they can be derived from the existing operand specifier and register-grouping control signals, which are already stored in the issue queues.
Figure \ref{fig:hazarding} depicts how instructions 3 and 4 in the example from Table \ref{tab:sboard} are resident in the issue queues and require no additional state beyond operand specifiers and an age tag.

Within the SIMD functional units, only the PWSb needs to be known, as register reads occur immediately after sequencing.
Since each micro-op in each pipeline stage in the SIMD functional units already needs to track a single destination register element group, the PWSb is trivially derived.
Figure \ref{fig:hazarding} depicts how instruction 0 is nearing completion in the arithmetic pipeline and requires no additional state to track the pending write hazard. 

\begin{figure}[h]
\centering
\includegraphics[width=0.9\linewidth]{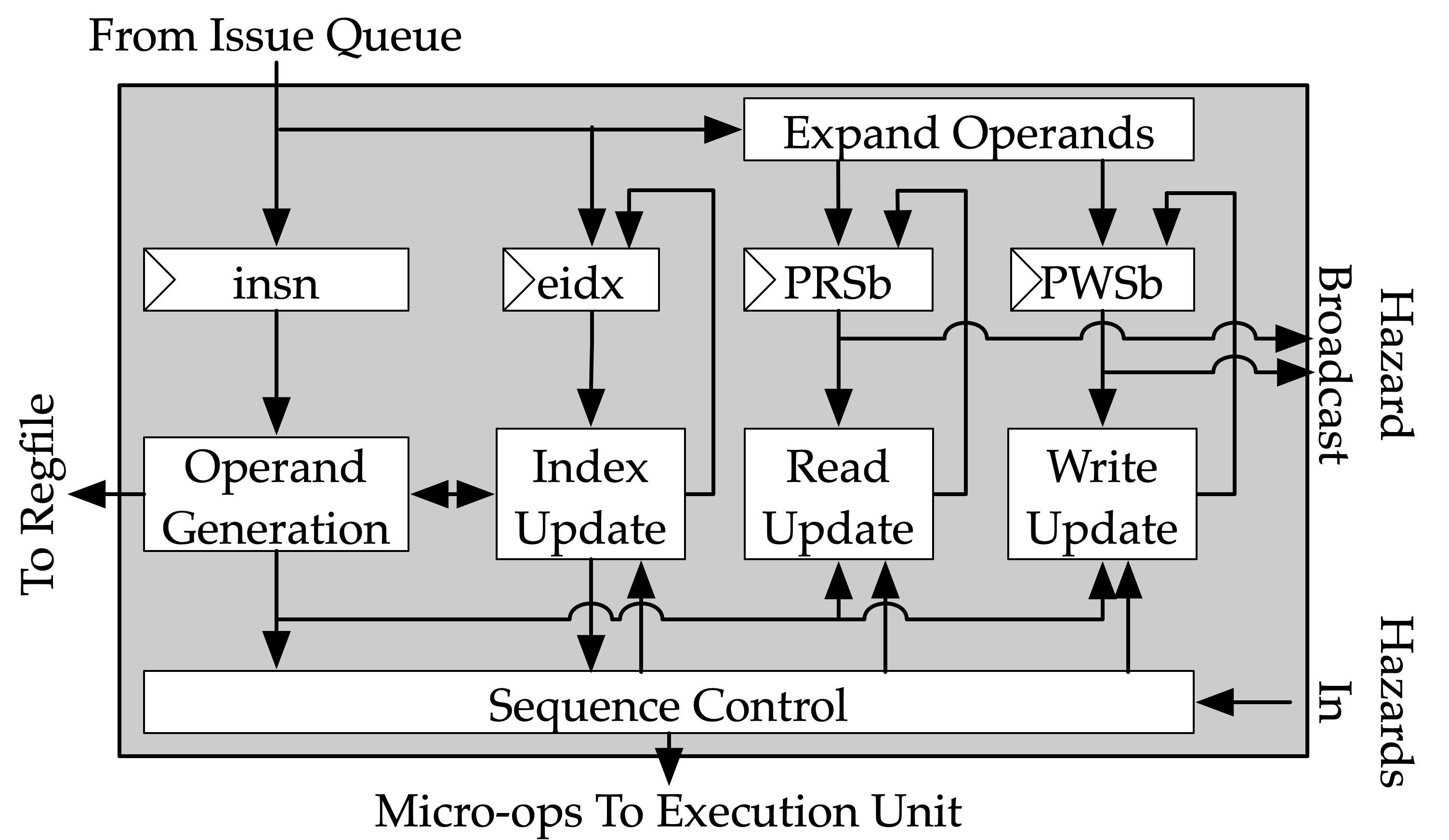}
\caption{Diagram of a vector instruction sequencer. The sequencer tracks precise PRSb and PWSb bit-vectors, updating them when it determines the next micro-op is free-of-hazards.}
\label{fig:sequencer}
\end{figure}

The only component in the microarchitecture which maintains precise element-group-granularity PRSb/PWSbs are the sequencers.
These scoreboards are aggressively cleared as micro-ops are sequenced.
Since the number of sequencers in Saturn is small (only 3 for load/store/arithmetic), the overhead of this state is tolerable.
Figure \ref{fig:hazarding} depicts how the load and arithmetic sequencers track per-operand scoreboards for the currently-sequenced instructions 1 and 2.
Figure \ref{fig:sequencer} depicts how the microarchitecture of the sequencer updates its internal PRSb and PWSb bit-vectors.

\subsubsection{Sequencing Algorithm}
At the sequencing stage, each sequencer must validate that the register reads and writes effected by the next micro-op are free of hazards against any older instructions or micro-ops within the OoO instruction window.
To do this, the microarchitecture must perform age disambiguation across the instructions within the OoO window to determine hazards from older instructions.
The PRSb/PWSbs from the older instructions can be bitwise OR'd together to form per-sequencer PRSb and PWSbs.

RAW, WAW, and WAR hazards can all be checked precisely using the element-group specifiers of the reads and write of the next micro-op the sequencer will issue.
The presence of a hazard stalls sequencing.
While the broadcasts of the PRSb and PWSbs may seem to be prohibitively expensive, several key properties enable optimizations that reduce the fan-in into each sequencer.

\begin{itemize}
\item \textbf{Issue-queues are in-order}, implying that the sequenced instruction will always be older than any other instructions from its parent issue queue. Only adjacent issue queues, sequencers, and pipelined functional units need to be checked for pending reads and writes.
\item \textbf{The number of pipelined functional units is low} since the load path is effectively a zero-deep pipeline.
Thus, the cost of interlocks on pending writes resident in functional units is limited.
\item \textbf{The OoO instruction window contains few instructions} across the shallow issue queues.
Unlike in scalar microarchitectures, where speculative execution requires a very deep instruction window, OoO execution here is used only to improve load-balancing across datapaths.
\item \textbf{Sequenced micro-ops will always be the oldest write to any element group,} eliminating the need to compare age tags against micro-ops in-flight in the functional units
\end{itemize}

Once a micro-op is cleared of hazards, it can be issued to register-read and the functional units.
Micro-ops issued by the sequencer are fire-and-forget, proceeding irrevocably through the pipelined functional units.

If the issued micro-op guarantees that the currently sequenced instruction will no longer read or write that element group again, as is the case for regular vector instructions, the sequencer can clear the accessed bit in its PRSb and/or PWSb, as show in Figure \ref{fig:sequencer}.
This enables cycle-granularity chaining across instructions with irregular access patterns into the vector register file.
Conversely, irregular vector instructions that do not read or write their operands in a static order can avoid clearing the sequencer scoreboards, disabling chaining from these instructions.

\section{Implementation}
Saturn is a Chisel RTL implementation of the complete RISC-V vector extension 1.0, supporting all application-profile features, including memory translation, precise traps, and irregular addressing modes.
Saturn is deeply parameterized and can target a wide range of short-vector design points, from small element-wise area-minimal vector units to wider, more performant, yet still compact DSP-core-like implementations, by varying the VLEN and DLEN parameters.


Saturn targets integration with a RISC-V dual-issue 6-stage host core, interfacing with the host core's existing TLB, CSR-file, and scalar load-store path.
Precise traps, virtual memory support, and the scalar-vector sequential memory ordering are all provided.
Additionally, parallel execution of scalar and vector memory operations is supported by providing a path for the scalar load-store path to access the vector load/store CAMs in parallel with the L1 cache access.
Memory ordering violations are caught before the commit stage of the host core.

The vector memory system bypasses the scalar L1 data cache, instead accessing a shared last-level-cache.
This enables efficient execution of kernels where scalar loads stream through elements resident in the L1, while vector loads and stores use a dedicated high-bandwidth path to shared system memory.

\section{Evaluation}

\begin{figure*}[!t]
\centering
\includegraphics[width=\linewidth]{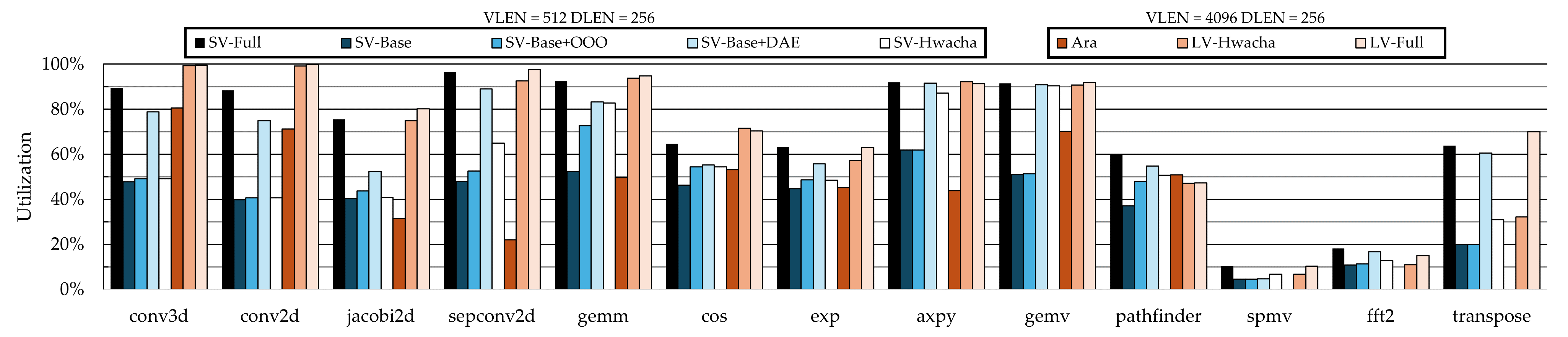}
\caption{Utilization across a variety of kernels. Comparison points include short-vector (SV) designs evaluated using Saturn, long-vector (LV) designs evaluated using Saturn, and Ara evaluated using its original RTL.}
\label{fig:perf}
\end{figure*}

We evaluate the performance, power, and area of Saturn in comparison to existing long-vector and short-vector implementations. 
For the area and frequency evaluations we use Cadence VLSI tools with a commercial 16nm process targeting 800 MHz at the the SS corner.


\begin{table}[b]
\centering
\caption{Workload Configurations}
\label{tab:bmarks}
\renewcommand*{\arraystretch}{1.1}
\begin{tabular}{|c|l|c|c|c|}
\hline
& \textbf{Benchmark} & \textbf{Problem Size} & \textbf{Datatype} & \textbf{LMUL} \\ \hline \hline
\multirow{5}{*}{\rotatebox[origin=c]{90}{\parbox[c]{5em}{\centering high-reuse}}} & \texttt{conv3d} & 112x112x7x7x3 & F64 & 2 \\
& \texttt{conv2d} & 112x112x7x7 & F64 & 2 \\
& \texttt{jacobi2d} & 130x130 & F64 & 4 \\ 
& \texttt{sepconv} & 119x119x3x3 & F32 & 4 \\
& \texttt{gemm} & 87x87 & F32 & 4 \\ \hline 
\multirow{4}{*}{\rotatebox[origin=c]{90}{\parbox[c]{1cm}{\centering no-reuse}}} & \texttt{cos} & 1024 & F32 & 4 \\
& \texttt{exp} & 1024 & F32 & 4 \\
& \texttt{axpy} & 30720 & F64 & 8 \\ 
& \texttt{gemv} & 128x128 & F32 & 8 \\ \hline
\multirow{4}{*}{\rotatebox[origin=c]{90}{\parbox[c]{4em}{\centering non elem.wise}}} & \texttt{pathfinder} & 64x1024 & I32 & 8 \\
& \texttt{spmv} & 128x128 60\% & F32 & 8 \\
& \texttt{fft2} & 1024 & F32 & 4 \\
& \texttt{transpose} & 180x180 & F32 & 1 \\\hline
\end{tabular}
\end{table}

\subsection{Performance}
\label{eval:perf}

We compare the utilization of Saturn across a range of benchmarks.
We report utilization as the main performance metric since all compared implementations share the same datapath throughput and memory bandwidth (both DLEN).
Utilization provides a portable and interpretable metric for understanding to what degree the various designs can saturate the fundamental structural resources of each machine.

Sufficiently large application sizes are chosen such that the long-vector microarchitectures are not penalized while keeping the working sets resident in the LLC and TLB.
The routines were implemented using C intrinsics or vector assembly, with the LMUL register grouping factor selected to maximize vector length while avoiding vector register spilling.
Table~\ref{tab:bmarks} depicts the problem sizes, datatypes, and LMUL for each benchmark.

For all comparison points, we use the same host dual-issue 6-stage core and cache-based memory systems.
This core can overlap RVV's \texttt{vsetvl} instructions alongside vector instructions, supporting 1 IPC issue into the vector unit for ideally scheduled code.
The outer memory system is configured with 4 banks of last-level cache, providing bandwidth of 256 bits/cycle and a total capacity of 512 KB, sufficient to hold the working set for all workloads after warm-up.
Access time to the cache is 4 cycles, but realistically degrades under load.

We evaluate a $\text{VLEN}=512, \text{DLEN}=256$ configuration of Saturn as \textbf{SV-Full}.
All comparison points described below, whether short-vector or long-vector, are configured with matching $\text{DLEN}=256$.

As a baseline, we configure a variant of Saturn without support for load-store decoupling and out-of-order issue.
This \textbf{SV-Base} variant is comparable to the Spatz~\cite{spatz} microarchitecture, since Spatz serializes execution through its global controller.
Among short-vector implementations, Spatz is the only comparable baseline, as it supports variable memory latency and does not require register renaming.
We could not compare directly against Spatz's implementation due to its partial support for RVV.
We also compare to \textbf{SV-Base+DAE} and \textbf{SV-Base+OOO} variants, enabling the decoupled load-store and out-of-order issue features of Saturn, respectively.

For long-vector microarchitectures, we simulate \textbf{Ara}'s RTL directly \cite{ara2} as a baseline implementation.
We use a 4-lane configuration of Ara to match the DLEN of the SV design and reproduce the integration approach in the Ara+CVA6 system with our dual-issue host core.

We also attempt a comparison with Hwacha's~\cite{hwacha_manual} scheduling mechanism by modelling 
Hwacha's fundamental behavior with modifications to Saturn's RTL.
Hwacha uses a central 8-entry master sequencer where complex instructions occupy multiple entries.
We evaluate this behavior for both short and long vector lengths, where \textbf{SV-Hwacha} uses $\text{VLEN}=512$ and \textbf{LV-Hwacha} uses $\text{VLEN}=4096$.

We additionally model a ``full-fury" no-restraints long-vector microarchitecture by setting $\text{VLEN}=4096$ without disabling any other features of Saturn.
This \textbf{LV-Full} configuration represents a hypothetical unrealistic long-vector design point.

Figure \ref{fig:perf} shows the utilizations across all the evaluation points.
The \textbf{SV-Base} configuration, lacking the ability to exploit memory-level parallelism or to dynamically load-balance across issue paths, suffers in all evaluated workloads.
While a decoupled load-store unit does improve performance in some simpler loops, like \texttt{axpy} or \texttt{gemv}, strict in-order sequencing results in frequent stalls in poorly load-balanced code.
Likewise, relaxing in-order sequencing alone is also insufficient, as a tightly coupled load-store path induces frequent RAW stalls anyways.
The \textbf{SV-Hwacha} configuration also underperforms, especially in kernels with complex instructions that would occupy multiple Hwacha sequencer slots.
Notably, \textbf{SV-Hwacha} suffers in the convolution kernels, where poorly scheduled loops greatly benefit when the microarchitecture can schedule across many inflight instructions.
The \textbf{SV-Full} results demonstrate that combining a DAE architecture with highly dynamic instruction scheduling and many inflight instructions is necessary for achieving near-peak $>90\%$ utilization across a wide range of kernels.

Compared to long-vector baselines, the \textbf{SV-Full} design exceeds the performance of the Ara implementation.
Even though the \textbf{SV-Hwacha} configuration underperformed, the \textbf{LV-Hwacha} design can still achieve $99\%$ utilization in some kernels, demonstrating that inefficient instruction scheduling can be offset by longer vector lengths.
Still, the \textbf{LV-Hwacha} design underperforms compared to \textbf{SV-Full} in \texttt{fft}, \texttt{spmv}, and \texttt{transpose}, demonstrating that long-vector-lengths cannot always compensate for scheduling inefficiencies.
Unsurprisingly, \textbf{LV-Full} achieves the highest utilization in almost all benchmarks. 
Section \ref{dse:chimes} presents further analysis on the implication of longer vector lengths in Saturn.

We compare Saturn's physical implementation results to reported numbers for prior vector units. 
As seen in Table \ref{tab:physical}, Saturn can achieve comparable frequencies to existing academic vector units.
Critical paths are in the SIMD datapath, which is not unique to Saturn's instruction scheduling.
Given that performance is equivalent to $(\text{utilization} \times \text{frequency})$ when the datapath throughputs are equivalent (DLEN is the same), we conclude that the \textbf{SV-Full} Saturn implementation achieves state-of-the-art performance among similarly configured academic vector units.

\begin{table}[h]
\centering
\caption{Physical Comparison of our Short Vectors Microarchitecture}
\label{tab:physical}
\renewcommand*{\arraystretch}{1.1}
\begin{tabular}{|wl{7.6em}|c|c|c|wc{2.6em}|wc{3.6em}|}
\hline
& \textbf{Ours} & \textbf{Spatz\textsubscript{2}} & \textbf{Ara} & \textbf{Hwacha} & \textbf{Vitruvius+} \\ \hline \hline
VLEN (bits) & 512 & 256 & 4096 & 4096 & 16384\\
DLEN (bits) & 256 & 64 & 256 & 256 & 512 \\ \hline
Process Node & 16nm & 22nm & 22nm & 16nm & 22nm\\
SS Freq. (MHz) & 800 & 485 & 950 & 800 & 1200\\
Area (kGE) & 1235 & 170\tablefootnote{Spatz\textsubscript{2} does not support floating-point or 64-bit operations.} & 2747 & 1118 & 6532 \\
Area (mm\textsuperscript{2}) & 0.41 & 0.14 & 0.55 & 0.38 & 1.3 \\ \hline
SGFLOPS & 12.5 & - & 15.2 & 12.5 & 18.6 \\
SGFLOPS/W (TT) & 121 & - & 90 & 57 & 47 \\ \hline
\end{tabular}
\end{table}

\subsection{Area}

Table \ref{tab:physical} provides a comparison of Saturn's area compared to numbers reported in publications of prior academic vector units.
These numbers capture only the area of the vector unit, ignoring any scalar core or memory-system components.
For prior work, kGE and mm\textsuperscript{2} numbers were extrapolated from published results and area breakdowns.
kGE provides a process-normalized approximation of area across process nodes.

The results show that Saturn's short-vector microarchitecture can yield a more area-efficient design than a few-lane implementation of a long-vector microarchitecture, suggesting better suitability for area-constrained deployment scenarios.

\begin{figure}[h]
\centering
\includegraphics[width=\linewidth]{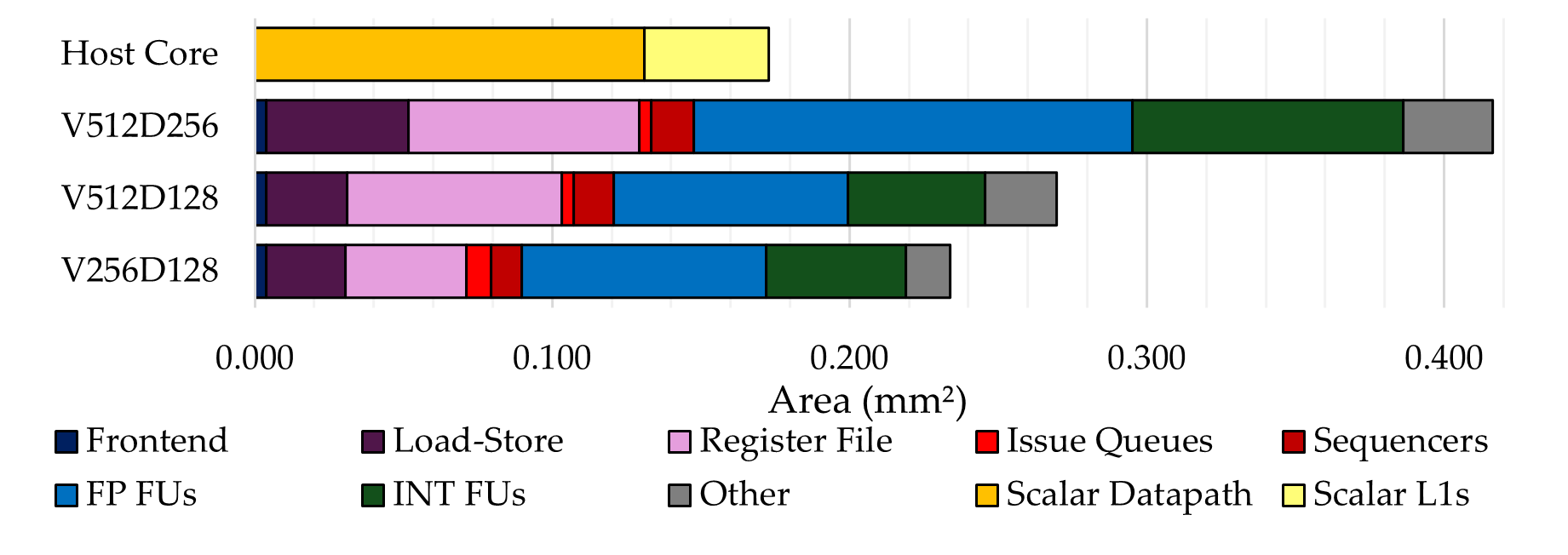}
\caption{Synthesized area breakdown across three configurations of Saturn alongside the breakdown of the host scalar core.}
\label{fig:area}
\end{figure}

\begin{figure}[h]
\centering  
\includegraphics[width=0.75\linewidth]{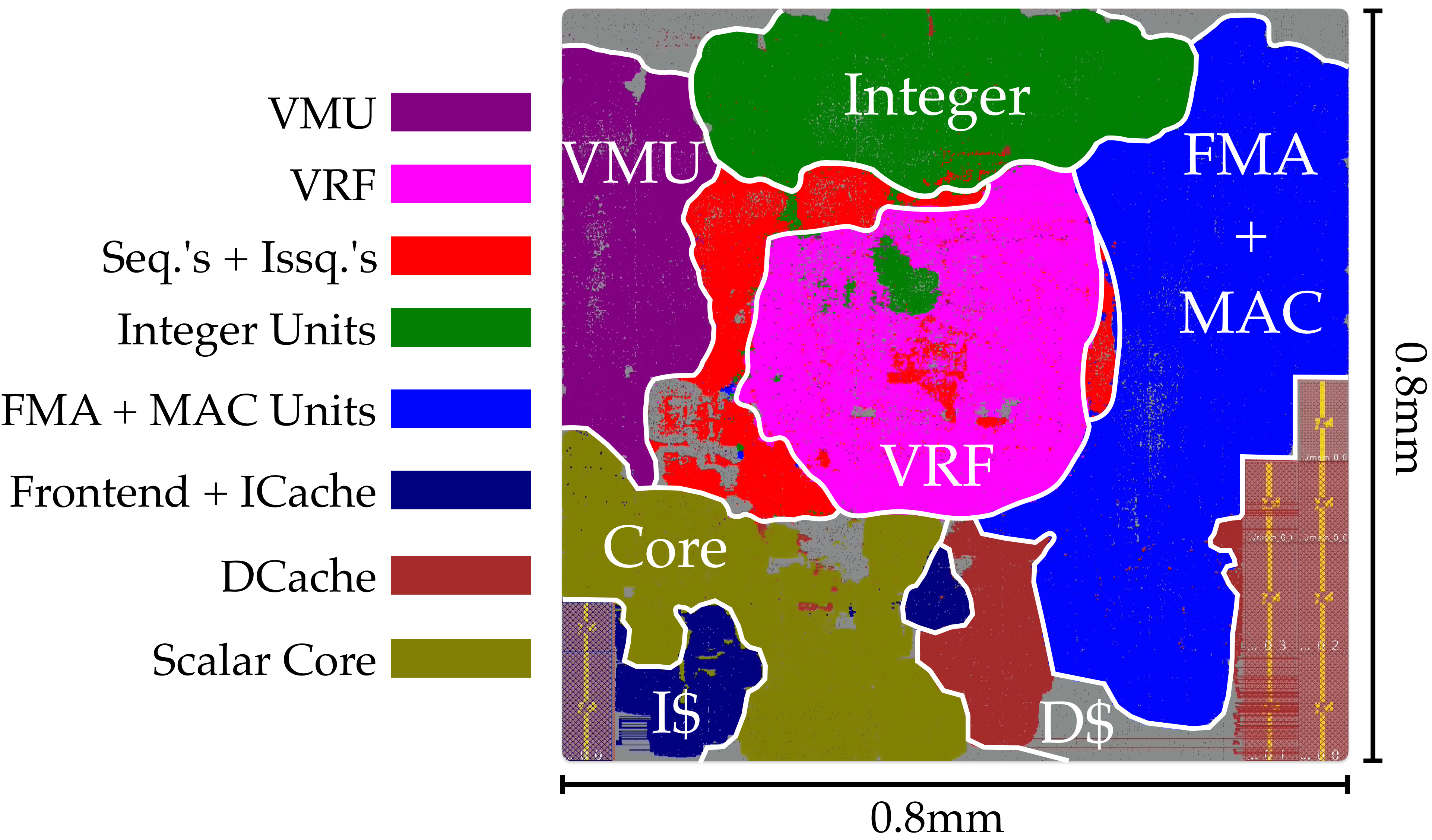}
\caption{Layout of the \textbf{SV-Full V512D256} implementation of Saturn, at 60\% density.}
\label{fig:layout}
\end{figure}

We additionally break down the area of the various components of Saturn across three machine configurations with varying VLEN and DLEN.
Figure \ref{fig:area} depicts this breakdown alongside the area of the host dual-issue core, while figure \ref{fig:layout} depicts a layout of one of the configurations.

As expected, the significant component of Saturn which scales linearly with architectural vector lengths is the vector register file.
The load-store unit and SIMD functional units, as the main datapath components, scale linearly with datapath width (DLEN).
The key components of Saturn's sequencing microarchitecture, the sequencers and issue queues, have consistently low area costs across all evaluated configurations.
The moderate scaling observed in the sequencers with higher DLEN reflects the area overhead of the accumulation registers, which are within the sequencer's module hierarchy.
The area of the frontend module for providing precise traps is marginal, demonstrating that supporting precise-traps in a decoupled vector unit has low cost.

The area results and layout show that Saturn enables compact area-efficient implementations built around a wide SIMD datapath.
The short-vector-specific components that enable efficient dynamic instruction scheduling are not costly compared to the necessary register file and SIMD functional units.






\subsection{Power}

\begin{figure}[h]
\centering
\includegraphics[width=\linewidth]{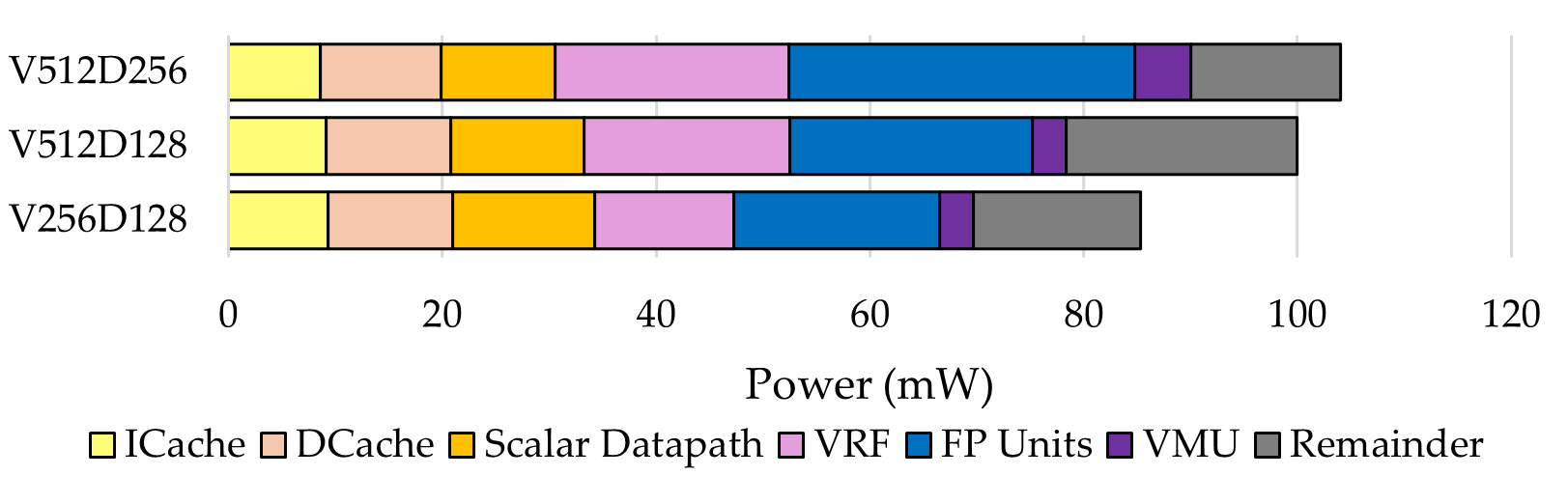}
\caption{Power consumption across three configurations of Saturn running a 64x64 SGEMM.}
\label{fig:power}
\end{figure}

We estimate the average power consumption of Saturn running a small 64x64 SGEMM kernel in Figure \ref{fig:power}. 
Power estimates are gathered post-synthesis at the TT corner and a clock frequency of 800MHz, annotated with waveform switching activities and post-PnR routing parasitics.

The power consumption of the VRF and vector FP units scales linearly with datapath width while remaining roughly constant with greater VLEN.
As noted in Table \ref{tab:physical}, the \textbf{SV-Full} design with $\text{VLEN}=512$, $\text{DLEN}=256$ design point achieves an efficiency of 112 SGFLOPS/W.
As noted in Table \ref{tab:physical}, the \textbf{SV-Full} design with $\text{VLEN}=512$, $\text{DLEN}=256$ design point achieves an efficiency of 121 SGFLOPS/W.
The power breakdown shows that the scalar core and caches consume a significant portion of each design's power when executing short-vector-length code.
Further optimizations to the scalar core, chiefly around reducing spurious ICache accesses within a low-IPC vector loop, could improve Saturn's power efficiency even further.





\section{Discussion}
In this section, we explore several of the key parameters of Saturn.
We discuss and evaluate Saturn's sensitivity to native chime length, issue queue depth, memory latency, and application vector length.

\newcommand{\x}{\phantom{x}}

\newcommand{\p}{\phantom{\%}}

\begin{table}[!t]
\centering
\caption{Percent Speedup with Increasing Chime Length (VLEN/DLEN) and Issue Queue Depth, with DLEN=256}
\label{tab:dseresults}
\renewcommand*{\arraystretch}{1.1}
\begin{tabular}{|wl{5.7em}||*{3}{wr{2.4em}|}*{3}{wr{2.3em}|}}
\hline
& \multicolumn{3}{c|}{\begin{tabular}{@{}c@{}}\textbf{Relative \% Speedup w.} \\ \textbf{Increasing VLEN/DLEN}\end{tabular}}
& \multicolumn{3}{c|}{\begin{tabular}{@{}c@{}}\textbf{Relative \% Speedup w.} \\ \textbf{Increasing IQ Depths}\end{tabular}} \\
\hline
\textbf{IQ Depths}  & 4 & 4 & 4 & 0$\rightarrow$1 & 1$\rightarrow$2 & 2$\rightarrow$4 \\
\textbf{VLEN/DLEN}  & 1$\rightarrow$2 & 2$\rightarrow$4 & 4$\rightarrow$8 & 2 & 2 & 2 \\
 \hline \hline
\multicolumn{1}{|l||}{\texttt{conv3d}}        &  57\% &  22\% &   1\% &       2\% &   3\% & 3\% \\
\multicolumn{1}{|l||}{\texttt{conv2d}}        &  61\p &  20\p &  -1\p &       3\p &   5\p & 3\p \\
\multicolumn{1}{|l||}{\texttt{jacobi2d}}      &  82\p &  -7\p &  -8\p &      12\p &  13\p & 6\p \\
\multicolumn{1}{|l||}{\texttt{sepconv}}       &  23\p &  -1\p &  -1\p &      20\p &   2\p & 0\p \\
\multicolumn{1}{|l||}{\texttt{gemm}}          &   3\p &   0\p &   5\p &       3\p &   1\p & 1\p \\ \hline
\multicolumn{1}{|l||}{\texttt{cos}}           &   2\% &   1\% &  -1\% &      11\% &   4\% & 4\% \\
\multicolumn{1}{|l||}{\texttt{exp}}           &  10\p &   4\p &   0\p &       2\p &  -1\p & 0\p \\
\multicolumn{1}{|l||}{\texttt{axpy}}          &   7\p &   4\p &   0\p &      -1\p &   0\p & 0\p \\
\multicolumn{1}{|l||}{\texttt{gemv}}          &   1\p &   2\p &   0\p &       1\p &   0\p & 0\p \\ \hline
\multicolumn{1}{|l||}{\texttt{pathfinder}}    &  57\% &   9\% &   1\% &      26\% &  -5\% & 3\% \\
\multicolumn{1}{|l||}{\texttt{spmv}}          &   6\p &  -6\p &  -3\p &      50\p &  19\p & 3\p \\
\multicolumn{1}{|l||}{\texttt{fft2}}          &   2\p &   1\p &  -2\p &       7\p &   2\p & 1\p \\
\multicolumn{1}{|l||}{\texttt{transpose}}     &  21\p &  31\p &  -7\p &       1\p &   1\p & 0\p \\ \hline
\end{tabular}
\end{table}

\subsection{Chime Length}
\label{dse:chimes}

The ``native" chime length in Saturn, for instructions with no register grouping, is given by the VLEN:DLEN ratio.
Increasing the native chime length reduces overall throughput pressure on many components in Saturn, including instruction fetch, load-store queue sizes, issue-queue sizes, and sequencer throughput.

However, a higher native chime length for a constant datapath linearly scales the size of the multi-ported register file.
The interlock logic in Saturn scales linearly with the chime length as well. 
We evaluate several configurations with varying VLEN:DLEN ratios, varying VLEN with fixed $\text{DLEN}=256$.
As we increment the VLEN:DLEN ratio, we record the percent improvement in performance for each kernel.
Table \ref{tab:dseresults} depicts the results of this comparison.

A native chime length of 1 imposes a very high performance burden, as this implementation effectively requires 1 IPC instruction issue when executing non-register-grouped code.
Moving a 2:1 ratio yields significant performance improvements across most kernels, as greater chime lengths diminish the performance impact of scalar stalls.
A 2:1 ratio is feasible with the precise scheduling mechanism and is the target design point for Saturn.

The effect is largely diminished at a 4:1 ratio, where only a few low-LMUL kernels see the benefit.
We additionally observe that some benchmarks actually exhibit performance \textit{degradation} at high chime lengths.
Analysis of these cases revealed that deep temporal execution diminishes the effectiveness of load balancing across issue paths.
These cases exposed a performance peculiarity specific to our evaluation system's memory system: the evaluated LLC achieves higher throughput with a dynamic mix of loads and stores, compared to separate load-intensive and store-intensive phases.

\subsection{Issue Queue Depth}

In design of scalar microarchitectures, deep issue queues are critical for maintaining high IPC, as the issue queues comprise the out-of-order execution window for the execution units.
However, Saturn's vector microarchitecture does not rely on the issue queue depth directly as a mechanism for maintaining functional unit utilization.

Rather, Saturn relies on the issue queues solely as a mechanism to enact load-balancing for suboptimal code sequences.
While optimally scheduled code might be perfectly balanced across all the sequencing paths, common vector code could benefit from some degree of per-sequencer queuing. 

Since Saturns' scheduling microarchitecture scale with the issue queue depths, shallow issue queues are especially desirable.
Thus, we evaluate the degree to which vector code sequences benefit from instruction queuing.
We compare variants of Saturn with issue queue sizes ranging from 0 to 4, recording the percent improvement in performance as we increment the queue depths.

The results in Table \ref{tab:dseresults} show that moving to single-entry issue queues immediately yields performance uplifts across many kernels.
Again, the effect diminishes rapidly, becoming insignificant for many kernels towards 4-deep issue queues.
Issue queue depths of 2-4 are likely the target design point for Saturn.
As shown in Figure \ref{fig:area}, the area cost of the issue queues is minimal, as Saturn does not impose substantial additional storage requirements for those queues. 

\subsection{Memory Latency}

\begin{figure}[h]
\centering
\includegraphics[width=\linewidth]{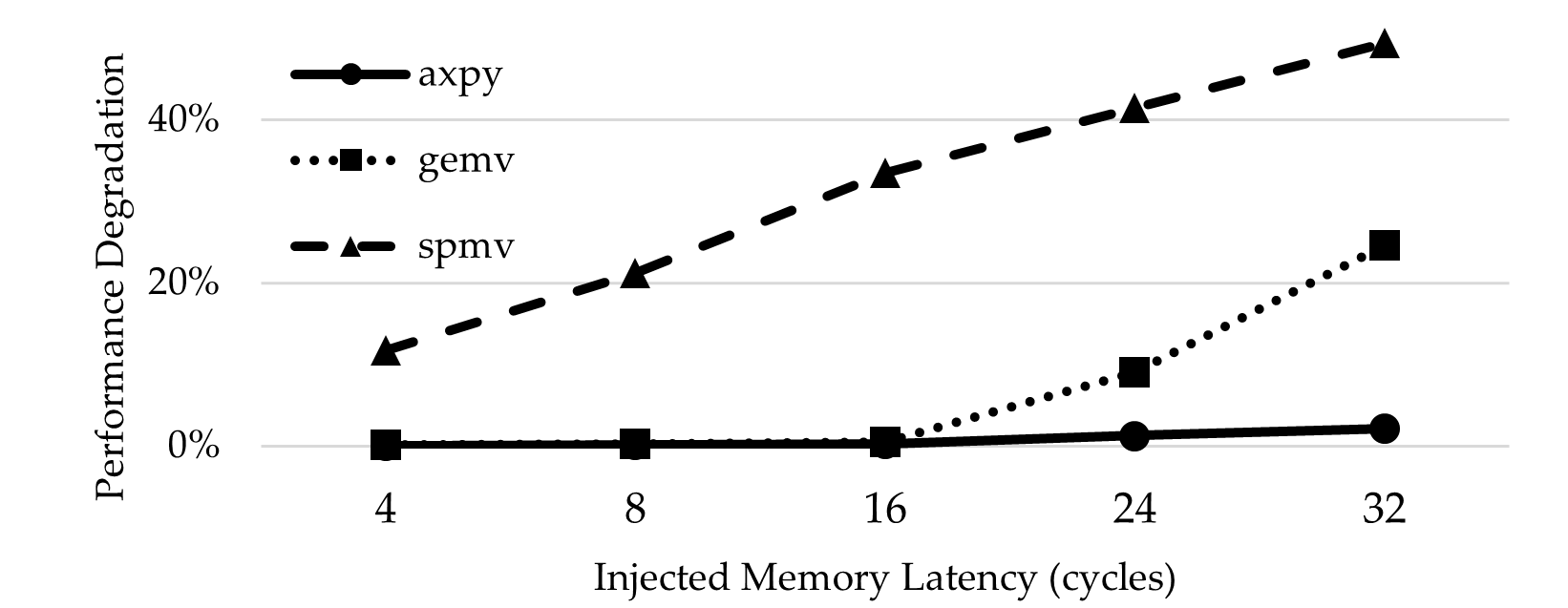}
\caption{Performance degradation with memory latency injection on top of the base latency to the last-level cache.}
\label{fig:latency}
\end{figure}

We evaluate Saturn's tolerance for memory latency through the DAE design paradigm of the vector load-store unit.
Unlike long-vector microarchitectures, a short-vector implementation cannot rely on hiding memory latency across long vector lengths.
The DAE microarchitecture enables run-ahead load request generation with minimal impact on the backend microarchitecture.
We add a latency-injection buffer between the load-store unit and the memory system and evaluate the performance degradation with increasing memory latency.

Figure \ref{fig:latency} shows that across a variety of memory-bound kernels, the DAE microarchitecture can effectively tolerate high memory latencies.
Notably, the SPMV kernel suffers more significantly from memory latency since the current frontend cracks page-crossing indexed loads into single-element operations, decreasing the effectiveness of the decoupling queues.
Reducing the degree of cracking performed under these cases or avoiding page-crossing indexed loads entirely would ameliorate this issue.

We can analyze the extent of tolerable memory latency by considering the decoupling and issue queue depths.
For a DAE microarchitecture, the maximum tolerable memory latency is determined by the depth of the queues between the access processor and the execute processor.

In Saturn, this includes both the explicit decoupling queue and the load issue queue.
Furthermore, a single vector instruction might encode multiple cycles of memory access.
Thus, the maximum tolerable memory latency is the number of instructions in the decoupling queue and load-issue-queue, multiplied by the maximum register grouping, multiplied by the native chime length.

In a $\text{VLEN}=512$ $\text{DLEN}=256$ machine with a 4-entry dispatch queue and a 4-entry load issue queue, this is 128 cycles of load latency in the optimal case, far more than would be necessary for a standard cached memory system.
As the decoupling queue plays no role in the instruction scheduling microarchitecture, its capacity can be increased at will when integrating Saturn into high-latency memory systems.

\subsection{Application Vector Length}

\begin{figure}[!h]
\centering
\includegraphics[width=0.9\linewidth]{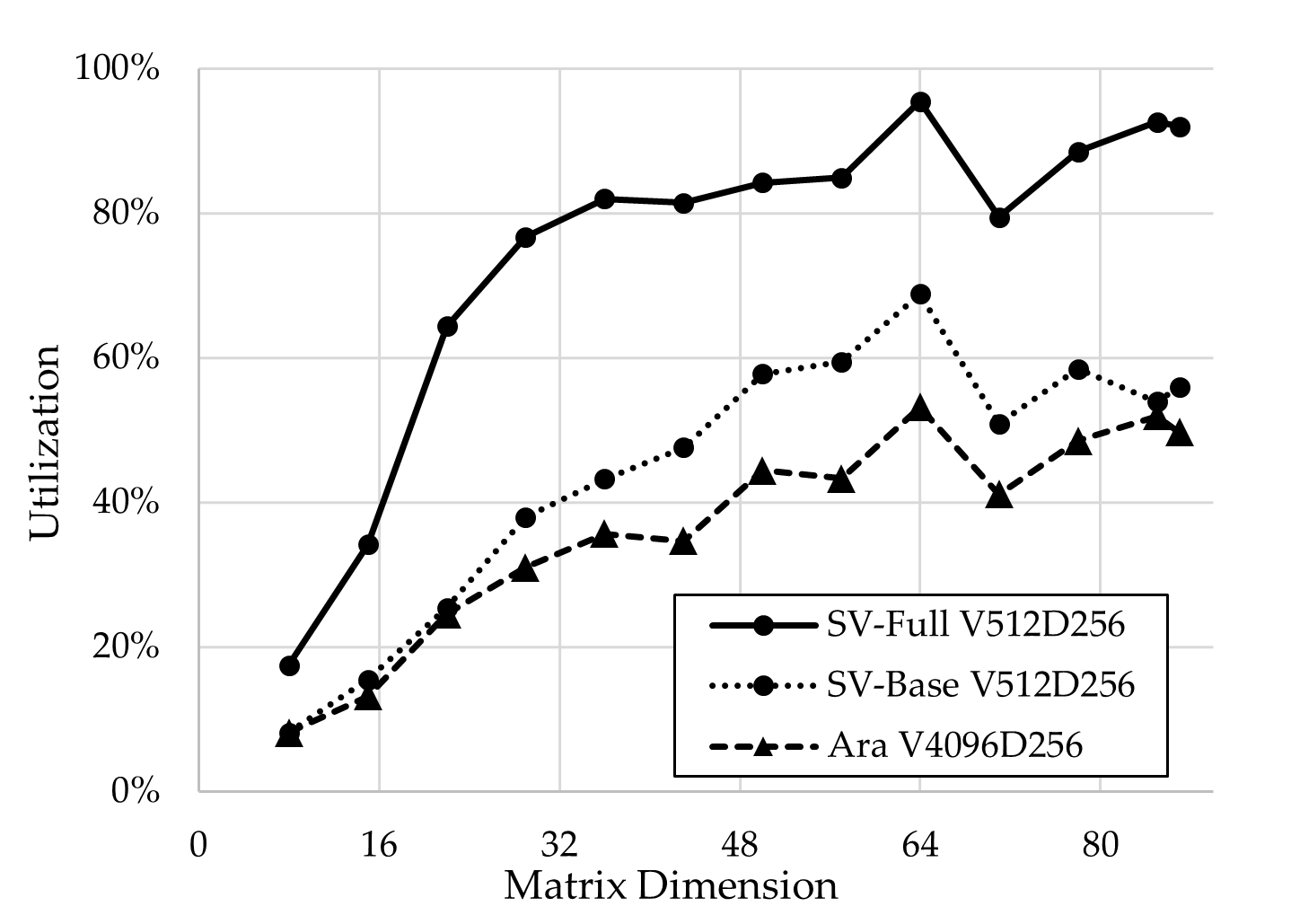}
\caption{Utilization of SGEMM with varying problem sizes.}
\label{fig:size}
\end{figure}

We sweep problem size for the compute bound floating-point matrix-multiplication algorithm and show the results in figure \ref{fig:size}.

With short vector lengths and shallower temporal execution, instruction throughput becomes more critical.
In the \textbf{SV-Base} configuration, inflexible scheduling reduces the effectiveness of the vector unit at draining instructions.
Both the \textbf{SV-Base} and \textbf{Ara} designs cannot reach their peak utilizations without a application vector length of 48.

In contrast, the \textbf{SV-Full} configuration can achieve near its peak utilization with a shorter application vector length of 32 elements.
Thus, Saturn's short-vector approach is more suitable for domains where application vector lengths are frequently short, as are in mobile, DSP, or embedded deployments.

\section{Related Work}

Existing work on vector microarchitectures have either required long-vector lengths, fixed low-latency memory systems, register renaming, imprecise traps, and/or a simpler vector ISA.
Saturn is the first to provide a full no-compromises short-vector implementation of a complete modern vector ISA.
Table \ref{tab:comparisons} summarizes the salient differences of Saturn against the most notable comparable academic and commercial vector implementations.

Recent academic interest in vector microarchitecture has been focused on long-vector machines for HPC applications.
These microarchitectures operate with fundamentally different constraints compared to short-vector units.
Deep temporal execution with low IPC and scalability towards many lanes are critical concerns for long-vectors, but do not reflect the requirements for short-vector units.

Among comparable work, Saturn is the first to demonstrate efficient execution of short vector lengths without requiring register renaming or a constrained memory system.
RISC-V\textsuperscript{2} and Ava require register-renaming, simplifying hazard determination at the expense of register-file size.
Saturn does not require register-renaming, but can still chain past all types of data hazards.
Spatz, Vicuna, and Torrent all assume a low-latency memory system.
Vicuna and Torrent further require a global memory stall to adapt to variable latency, a significant limitation.
Saturn supports dynamic instruction scheduling and chaining in the presence of long and variable-latency memory systems.

Many commercial vector implementations, including the Xuantie 910\cite{c910}, SiFive P-series\cite{sifivep}, Ventana Veyron\cite{ventana}, Semidynamics Atrevido~\cite{semidynamics}, modern ARMv9 cores\cite{armv9}, and Fujitsu A64FX\cite{a64fx}, integrate their vector units deeply within the out-of-order core microarchitecture.
Compared to these implementations, Saturn does not rely on general-purpose-optimized capabilities of the host core, such as aggressive superscalar instruction fetch, arbitrary out-of-order execution, or register renaming.

\begin{table}[!t]
\centering
\caption{Comparing Related Vector Microarchitectures}
\label{tab:comparisons}
\renewcommand*{\arraystretch}{1.05}
\begin{tabular}{|c|l|c|c|cc|cc|}
\hline
&
& \rotatebox[origin=c]{90}{\textbf{Classification}}
& \rotatebox[origin=c]{90}{\textbf{VLEN (Bits)}}
& \rotatebox[origin=c]{90}{\textbf{No Renaming}}
& \rotatebox[origin=c]{90}{\textbf{Var. Mem. Lat.}}
& \rotatebox[origin=c]{90}{\textbf{Prec. Traps}}
& \rotatebox[origin=c]{90}{\textbf{RVV 1.0}}
\\ \hline \hline
\multirow{9}{*}{\rotatebox[origin=c]{90}{\parbox[c]{5em}{\centering \textbf{Academic}}}} 
& \textbf{This work} & Short & 128-1024 & \cmark & \cmark & \cmark & Full \\ \cline{2-8}
& Torrent~\cite{t0} & Long & 1024 & \cmark & \xmark & \xmark & \xmark \\ 
& OOOVA~\cite{ooova} & Long & 8192 & \xmark & \cmark & \cmark & \xmark \\
& Hwacha~\cite{hwacha_manual} & Long & 4096 & \cmark & \cmark & \xmark & \xmark \\
& Ara~\cite{ara2} & Long & 4096 & \cmark & \cmark & \xmark & Partial \\
& Vitruvius+~\cite{minervini_vitruvius_2023} & Long & 16384 & \xmark & \cmark & \xmark & Partial \\ 
& AVA~\cite{adaptable_regfile} & Long & 1024 & \xmark & \cmark & \xmark & Partial \\ \cline{2-8}
& Spatz~\cite{spatz} & Short & 512 & \cmark & \xmark & \xmark & Partial \\
& Vicuna~\cite{vicuna} & Short & 512 & \cmark & \xmark & \xmark & Partial \\
& RISC-V\textsuperscript{2}~\cite{patsidis_risc-v2_2020} & Short & 128 & \xmark & \cmark & \xmark & Partial \\ \hline
\multirow{5}{*}{\rotatebox[origin=c]{90}{\parbox[c]{5em}{\centering \textbf{Industrial}}}} 
& NEC Aurora~\cite{nec_aurora} & Long & 16384 & \xmark & \cmark & \xmark & \xmark \\
& A64FX~\cite{a64fx} & Short & 512 & \xmark & \cmark & \cmark & 
\xmark \\
& P-series~\cite{sifivep} & Short & 256 & \xmark & \cmark & \cmark &  Full \\
& X-series~\cite{sifivex} & Short & 512 & \cmark & \cmark & \cmark &  Full \\
& NX27V~\cite{nx27v} & Short & 512 & \cmark & \cmark & \cmark & Full \\ \hline
\end{tabular}
\end{table}

The Andes NX27V~\cite{nx27v} and SiFive X~\cite{sifivex} series processors are commercial RVV implementations with similar short vector lengths and compact, unified datapaths.
While the details of these commercial implementations are unknown, we observe several key differences with Saturn.
The NX27V seems to implement a unified physical scoreboard, while Saturn's scheduler distributes the scoreboard across sequencers.
The X-series feature a unified scalar and vector load-store unit, while Saturn's load-store unit is vector-specialized and separate from the scalar access path.


\section{Conclusion}

We propose, implement, and evaluate a short-vectors microarchitecture that addresses the requirements for general-purpose data-parallel compute in compact efficient cores.
Our implementation can achieve competitive performance compared to aggressive long-vector microarchitectures, while also fulfilling all the necessary architectural requirements in modern scalable vector ISAs.

\section{Acknowledegments}
Research was partially funded by SLICE Lab industrial sponsors and affiliates, and by the NSF CCRI ENS Chipyard Award \#2016662.
Any opinions, findings, conclusions or recommendations expressed in this material are those of the author(s) and do not necessarily reflect the views of the National Science Foundation.
We thank Kevin He, Nico Casteneda, Mihai Tudor, and Vikram Jain for their work leading tapeouts which have integrated early versions of Saturn.
We also thank Albert Ou, Colin Schmidt, Andrew Waterman, and Chris Batten for insightful conversations on vector microarchitecture.

\bibliographystyle{IEEEtranS}
\bibliography{refs}

\end{document}